# Crashworthiness design of 3D lattice-structure filled thin-walled tubes based on data mining

Jiyuan Lv[1], Zhonghao Bai[1], Xianping Du[2], Feng Zhu[3], Clifford C. Chou[4], Binhui Jiang[1*], Shiwei Xu[1*]

[1] The State Key Laboratory of Advanced Design and Manufacturing for Vehicle Body, Hunan University, Hunan, 410082, China

[2] Department of Mechanical Engineering, Embry-Riddle Aeronautical University, U.S.A.

[3] Hopkins Extreme Materials Institute, The Johns Hopkins University, U.S.A.

[4] Bioengineering Center, Wayne State University, MI 48201, USA

*Corresponding author:

Tel.: +86-13787054434

Fax: +86-731-88823715

Email address: jjhhzz123@163.com (Binhui Jiang) and xushiwei@hnu.edu.cn (Shiwei Xu)

The word count for the text without abstract and references is 4727. And the word count for the abstract is 311.

The number of figures and tables are 12 and 6, respectively.

---

* Corresponding author: jjhhzz123@163.com (Binhui Jiang) and xushiwei@hnu.edu.cn (Shiwei Xu).

# Crashworthiness design of 3D lattice-structure filled thin-walled tubes based on data mining

Jiyuan Lv[1], Zhonghao Bai[1], Xianping Du[2], Feng Zhu[3], Clifford C. Chou[4], Binhui Jiang[1*], Shiwei Xu[1*]

[1] State Key Lab of Advanced Design and Manufacturing for Vehicle Body, Hunan University, China
[2] Department of Mechanical Engineering, Embry-Riddle Aeronautical University, U.S.A.
[3] Hopkins Extreme Materials Institute, The Johns Hopkins University, U.S.A.
[4] Bioengineering Center, Wayne State University, U.S.A.

**Abstract**: Lattice structures and thin-walled tubes are two types of energy-absorbers widely studied and applied in engineering practice. In this study, a new type of lattice-structure filled thin-walled tube (LFT) was proposed by combining these two energy absorbers. In this new type of LFT, a BCC-Z (Body-centered cubic unit cell with vertical struts) lattice structure was filled into a square thin-walled tube. Then using data mining, a 3-D geometric design with five design variables, consisting of the number of layers in the longitudinal direction $n$, number of cells in the transverse direction $m$, rod diameter $d$, tube thickness $t$, and height difference between tube and lattice structure $h$, was conducted on this new LFT. Using Latin Hypercubic sampling algorithm, 150 design cases were generated. Numerical models were then developed to simulate their crush behavior, and the simulation dataset was used for data mining. The results showed that (1) Filling the BBC-Z lattice structure into a thin-walled tube can significantly improve the energy absorption (EA) capacity of the structure. (2) SEA (Specific Energy Absorption) of LFT significantly increased by increasing the rod diameter $d$, number of cells in the transverse direction $m$, and number of layers in longitudinal direction $n$. (3) The decision trees generated in the data mining process indicated that the rod diameter $d$ of lattice structure is the key design variable that has most significant impact on EA, followed by $m$ and $n$. (4) The design rules to build LFTs with high EA efficiency (SEA≥16 kJ/kg and CFE (Crush Force Efficiency)≥45%), high total EA (SEA≥16 kJ/kg and EA≥6 kJ), and lightweight (SEA≥16 kJ/kg and Mass≤0.45 kg) were obtained from decision trees. The ideal configurations of LFT corresponding to these three objectives are: d>2 mm, n>2 and m>3 for high EA efficiency; d>2 mm, n>2 and m>3 for high total EA; and d>2 mm, n>2, m≤4 and t≤1.7 mm for lightweight.

[*] Corresponding author: jjhhzz123@163.com (Binhui Jiang) and xushiwei@hnu.edu.cn (Shiwei Xu).

## 1. Introduction

As a high performance and widely used energy absorber in aerospace, automotive, and other fields [1,2], thin-walled tubes have been extensively studied, e.g. the effects of material type (steel, aluminum alloy, composite material, etc.) on their energy absorption (EA) characteristics; influence of **2-D cross-sectional shapes** (circular, square, hexagon, octagon, mixed shapes, etc.) [3-6] or longitudinal curvature change (ripple, sine-curve, hierarchical gradient, etc.) [7-11]. In addition, it was found that filling thin-walled tube with low-density materials or structures (such as foam, honeycomb structure, etc.) could greatly improve the EA capacity [12-15].

Due to the lightweight, high porosity, specific strength, and EA efficiency, lattice structures can also be used as the filler to enhance the EA capacity of thin-walled tubes. For example, Kim et al. (2019)[16] reported that a thin-walled tube filled by multiple layer pyramidal truss cores could absorb more energy than a hollow thin-walled tube. Cetin and Baykasoğlu (2019)[17] studied another lattice filled tube and obtained similar results. **In addition to the improvement of the energy absorption performance of the thin-walled tube by filling the lattice-structure, the 3-D gradient change and design for the mechanical properties of thin-walled tube can be realized by adjusting the filling density of the lattice-structure.** However, due to the structural complexity and difficulty in fabrication, lattice structures were not widely found in engineering applications. In recent years, with the development of 3D printing technology, manufacturing of lattice structures has become much easier [18-20]. Many studies have been conducted on the lattice structures [16,21-25] and much wider applications of lattice-structure filled thin-walled tube (LFT) in engineering aspects also become possible. The studies of lattice structures appeared in the literature [26-30] mainly include BCC (Body-centered cubic), BCC-Z (Body-centered cubic with vertical struts), and $F_2BCC$ (Face centered cubic in combination with BCC). Compared to BCC and $F_2BCC$, vertical bars (or struts) in BCC-Z can provide more supports during the compression[28,29]. Gümrük et al. (2013)[30] also reported that BCC-Z exhibited a better crushing performance compared to BCC and $F_2BCC$.

From the above review, it has been found that much less studies have been conducted on LFT than lattice structures alone. For example, no study has considered the effects of type, density, strength etc. of filling lattice structure on the EA characteristics of LFT. Too many parameters affecting the EA characteristics of LFT (e.g. the geometric or material parameters of tube and lattice structure) are major challenges in the traditional parametric study or optimal design of LFT. Since the traditional parametric study is hard to clarify the interrelationship between high-dimensional and intricate parameters [31,32], Such traditional approach is not efficient to analyze the complex structures. Additionally, in traditional structural optimal design, optimum is often obtained from a large number of simulations based on pre-defined optimization goals. It is quite often that such globally optimal design is only a single data point in the design space, and very difficult to achieve in reality due to the uncertainty in design and manufacturing. Therefore, people are more interested in obtaining a "group" of good designs which can satisfy the design objectives, rather than a single global optimum.

To address these issues, a new structural design methodology based on data mining was proposed by Zhu et al. (2016)[33] ,[34-36]. In this new approach, the decision tree algorithm, one of data mining methods, was used to mine the dataset of design results to obtain the interrelationships between design parameters and effect of these parameters on the structural responses. An interpretable tree type diagram was used to describe the design rules. In the decision tree, design variables with their value ranges are chained with the corresponding design result classification. Therefore, if the value ranges of parameters are determined in the simulations, the corresponding design results can be predicted by this decision tree. On the other hand, if the design requirements (classification) of results are identified, the value ranges of parameters can be determined by the decision tree. Due to these features, the decision tree method has been gradually extended from the fields of data analysis, such as finance, marketing and logistics, to solve various complex engineering problems including structural design, vehicle safety, and injury biomechanics[37-42]. In particular, **the decision tree in structural design can be used to derive design rules which can rapidly generate one or more sets of good design without inferior design regions in the whole design space**.

In this paper, the new design methodology based on data mining is used to study and design LFT. In this LFT, the thin-walled tube with constant length, width and height is filled with the BCC-Z lattice structure which has various rod diameters, longitudinal layers, and transverse distribution. Meanwhile, the effects of tube thickness and height difference between tube and lattice structure on the EA characteristics of LFT are also considered.

## 2. Material and Methods

### 2.1. Lattice-structure filled thin-walled tube (LFT)

Figure 1a shows the single unit of BCC-Z lattice structure used in this study described by three parameters, i.e. diameter $d$, length $l$, and horizontal inclination $\omega$ of rod. The parameters of square tube include the constant length\width $a=75$ mm, height $H=200$ mm, and the gap between the tube and lattice structure $S=1$ mm, tube thickness $t$ and height difference $h$ between the tube and lattice, as shown in Figure 1b. The density of filled BCC-Z is controlled by the number of layers in the longitudinal direction $n$ and number of cells in the transverse direction $m$. In Figure 1, $n$ is 4 and $m$ is 2. The 3-D parametric design of LFT is shown in Figure 1c.

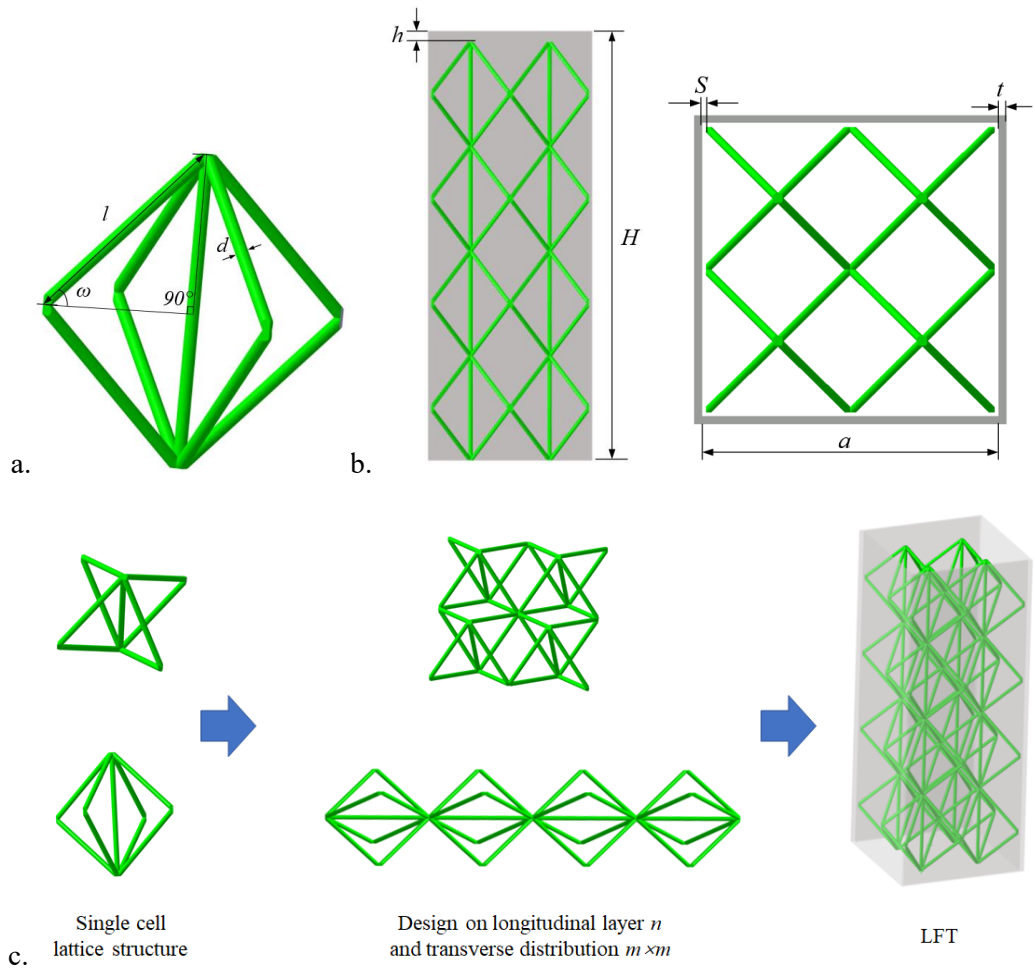

**Figure 1.** The schematic diagram for LFT design: (a). A single unit of BCC-Z, (b). A thin-walled tube fully filled by BCC-Z lattice, (c). Isometric view of the whole LFT

In order to fully fill the tube, length $l$ and horizontal inclination $\omega$ of rod were considered as the functions of $n$, $m$, and $h$ which are expressed in Equations (1) and (2), respectively. Therefore, there are five (5) independent design variables in LFT design, which are $n$, $m$, $d$, $t$, and $h$. Their value ranges are given in Table 1.

$$\omega = f(n, m, h) = \tan^{-1}((\frac{H-h}{2n})/(\frac{a-2s}{\sqrt{2}m})) = \tan^{-1}(\frac{\sqrt{2}m(200-h)}{146n}) \quad (1)$$

$$l = f(n, m, h) = (\frac{H-h}{2n})/\sin\omega = (\frac{200-h}{2n})/\sin\frac{\sqrt{2}m(200-h)}{146n} \quad (2)$$

where $a=75\ mm$ and height $H=200mm$ are substituted.

Table 1. Ranges of five (5) independent design variables in LFT design

| Variables | $n$ | $m$ | $d$/mm | $t$/mm | $h$/mm |
|---|---|---|---|---|---|
| Value range | 2~6 | 2~5 | 1~3 | 0.8~2 | 0~5 |

**2.2. Crashworthiness performance indices**

Several indices related to structural crashworthiness[3] were used to evaluate the EA characteristics and they are introduced as follows.

The total energy absorption (*TEA*) by LFT during the axial crush is expressed as

$$TEA = \int_0^z F(x)dx \quad (3)$$

where $F(x)$ is the transient compression force and z is the crush distance.

Specific EA is defined as the energy absorption per unit mass and written as [43]

$$SEA = \frac{TEA}{M} \quad (4)$$

where M is the total mass of LFT.

The mean crush force is related to the total energy absorption and calculated as

$$P_m = \frac{TEA}{z} = \frac{\int_0^z F(x)dx}{z} \quad (5)$$

Crush force efficiency (*CFE*) is the ratio between the mean crush force and peak crush force *PCF* and expressed in the form[3]

$$CFE = \frac{P_m}{PCF} \quad (6)$$

**2.3. Finite element modelling and validation**

A finite element (FE) model of LFT was developed using the explicit nonlinear FEA code LS-DYNA (version 971, LSTC, Livermore, CA) to study the EA characteristics of LFT under the axial loading. In this FE model, the LFT was located between two rigid plates. One plate was fixed to support the LFT at the bottom, the other one moved downwards at a constant velocity

*v=10 m/s* to crush the LFT (as shown in Figure 2). Four-node Belytschko-Tsay thin shell elements with 3 integration points through the thickness and 2 mm mesh size were applied to model the tube, and solid elements with 0.5~1.5 mm mesh size were used for the lattice structure. The *CONTACT_AUTOMATIC_SURFACE_TO_SUFRACE algorithm in LS-DYNA was used to calculate the contact between the rigid plates and LFT, and the *CONTACT_AUTOMATIC_SINGLE_SURFACE was also defined for the LFT to prevent any self-penetration. Static and dynamic coefficient of friction coefficients were defined as 0.3 and 0.2, respectively.

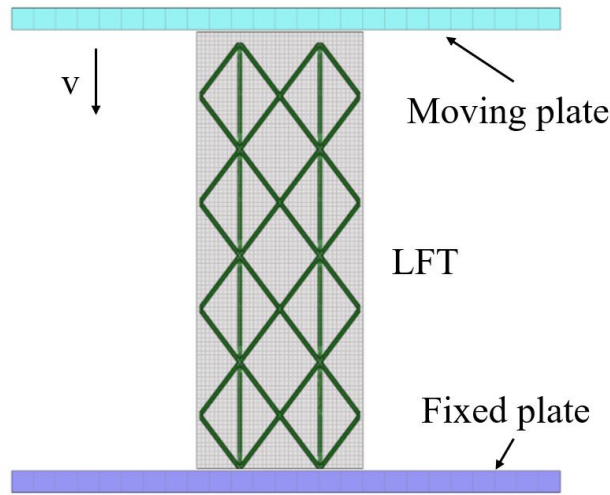

**Figure 2.** The FE model of LFT to simulate its crushing response

Aluminum alloy Al6063-T5 is commonly used to produce the EA tubes on vehicles. So, it was also employed in this study for the thin-walled tube. Considering that the lattice structure may be fabricated with 3D printers in future applications, it was assumed to be made of aluminum alloy AlSi10Mg, one of the metal materials widely used in 3D printing technology. Both materials (Al6063-T5 and AlSi10Mg) were simulated by a piecewise material constitutive law, known as *MAT_123 or *MAT_MODIFIED_PIECEWISE_LINEAR_PLASTICITY in LS-DYNA. The detailed material parameters are listed in Table 2. Their typical tensile stress-strain curves are shown in Figure 3.

**Table 2.** The major material parameters for the two aluminum alloys[17]

| Aluminum alloys | E (GPa) | Yield stress (MPa) | Density (kg/m$^3$) | Poisson's ratio |
| --- | --- | --- | --- | --- |
| Al6063-T5 | 68.2 | 187 | 2700 | 0.33 |
| AlSi10Mg | 69.3 | 162 | 2670 | 0.3 |

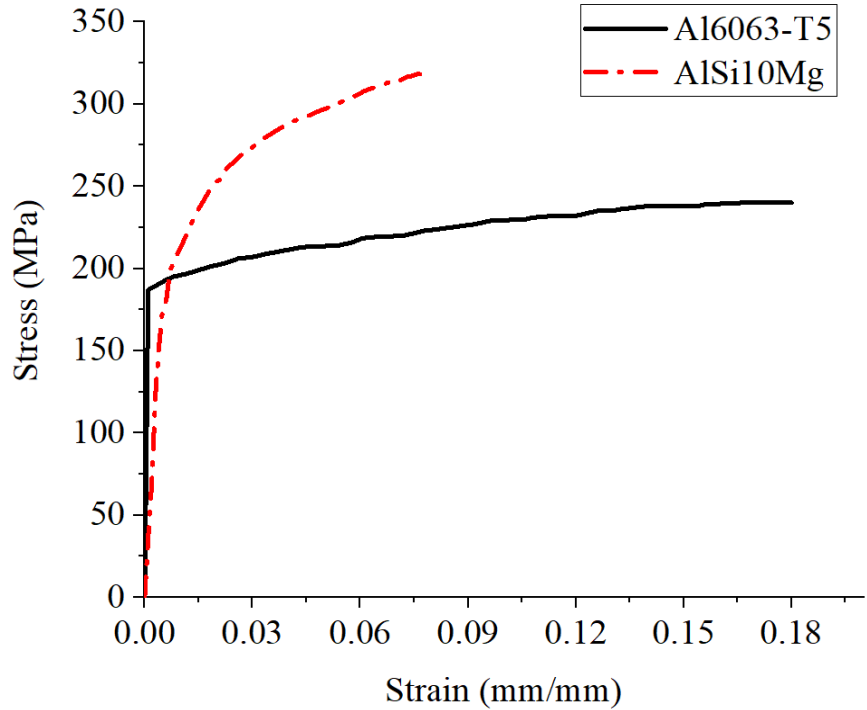

**Figure 3.** Tensile stress-strain curves of the two aluminum alloys [17]

The experimental data reported by Cetin and Baykasoğlu (2019)[17] were used to validate the FE model. In their study, a type of LFT filled BCC lattice structure with *m=1* and *n=4* was tested under axial compressive loads. The tube and lattice structure of tested LFTs were also made by Aluminum ally Al6063-T5 and AlSi10Mg, respectively. The dimensions of the FE model were identical to those of the specimens. The force-displacement curves obtained from test by Cetin and Baykasoğlu (2019)[17] and predicted simulation in this study are compared in Figure 4a. It was found that the trend of the simulation curve was basically consistent with that of the experimental curve. The initial peak force predicted by simulation was about 26.1 kN, which was lower than that of test (26.2 kN) by 0.23%. The mean crush force predicted by simulation was about 20.9 kN which was higher than that of test (19.9 kN) by about 5.1%. The deformation patterns of LFT predicted by simulation and observed in the tests are compared in Figure 4b. A good agreement has been demonstrated. Therefore, the performance of the FE model for LFT has been verified and can be used in the subsequent analysis and design.

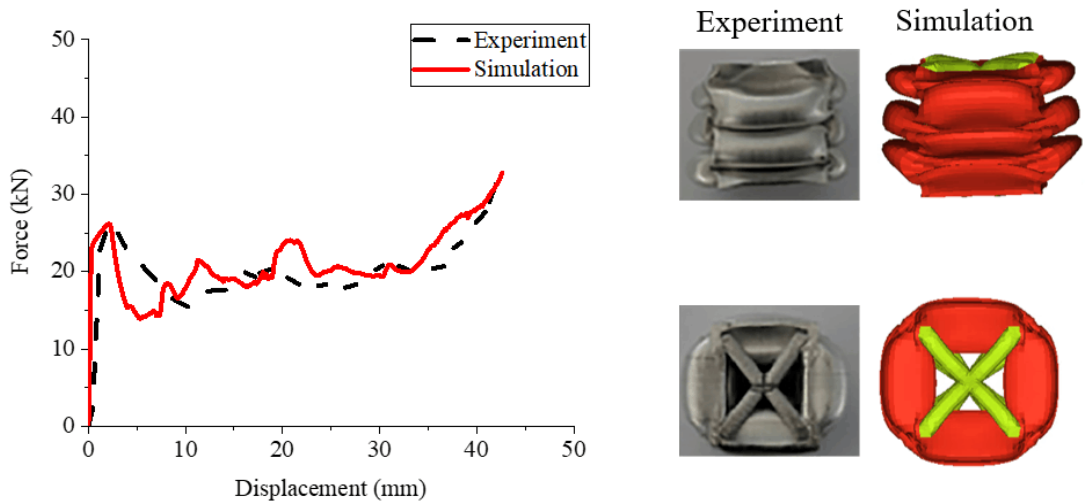

| a. Force-displacement curves comparison | b. Deformation patterns comparison |

**Figure 4.** A comparison of the results of simulation and test[17]

**2.4. Data mining-based design methodology**

Data mining is a process of acquiring knowledge and rules from a dataset through machine learning algorithms. The decision tree algorithm is one of the common machine learning algorithms [33-35], where the rules or relationships mined are described in the form of a tree-like diagram. The flow chart of design method for LFT based on data mining is shown in Figure 5. The whole design process goes through the following five steps:

- Step 1. the LFT is parameterized first and the design space should be defined for each design vairables.
- Step 2. A large number of design alternatives are generated within the design space, and each one is termed a sample in design of experiments (DOE) and used to create the corresponding FE model of LFT.
- Step 3. A dataset used for data mining is generated from simulations.
- Step 4. Descision trees are obtained through mining the simulation dataset.
- Step 5. The design rules are derived from the descision tree and then validated using simulation results.

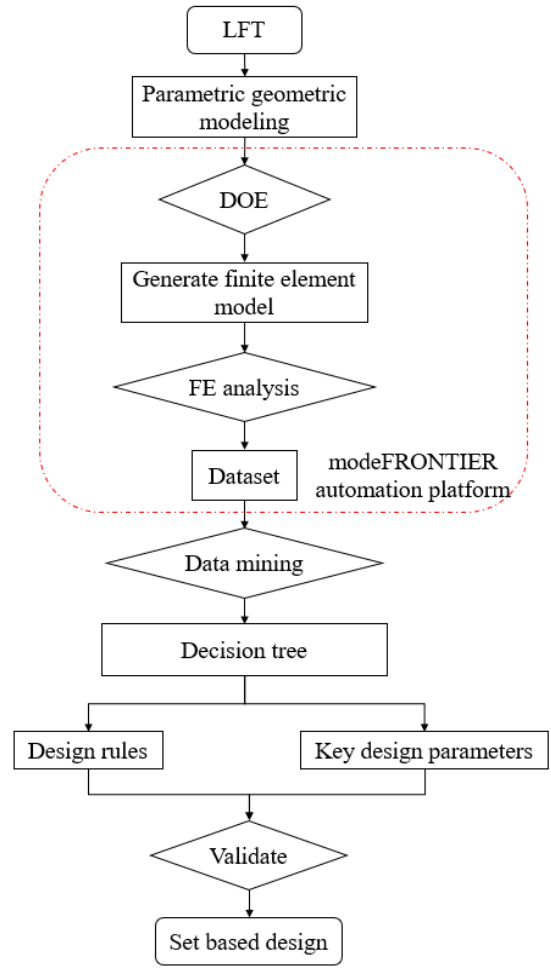

**Figure 5.** The flow chart of design method for LFT based on data mining

In order to build the simulation dataset, a large number of simulations should be conducted. An automation platform was developed with modeFRONTIER (2016, ESTECO, Trieste, Italy) to automatically complete the modelling, simulation and dataset construction. modeFRONTIER can integrate and call CAD, meshing and simulation software to implement each step without much manual intervention [44]. In this automation platform, Catia (V5, Dassault, Courbevoie, France) was used to generate geometric models. Meshing and simulation analyses were completed in Hypermesh (2017, Altair, Troy, MI, USA) and LS-DYNA, respectively. MATLAB (2017, MathWorks, USA) was then used to process the simulation results and calculate and collect *TEA*, *SEA*, $P_m$ and *CFE* data for each design alternative. After the dataset was ready, the decision tree was established using WEKA (V3.8.3, Waikato University, New Zealand). The decision tree is a tree like diagram consisting of one root node, several leaf nodes, non-leaf nodes and branches. When it is used in the structural design, non-leaf nodes represent the design variables; branches are used to specify the value ranges of the design variables, and the leaf nodes serve as the labels for the classification in terms of structural EA characteristics. The effects of design variables on the structural EA characteristics can be interpreted from the root node which represents the most influential variable: a branch under each decision node shows the value range of a design variable in a certain design. A path from the root to a leaf node is essentially a decision-making process. Therefore, a decision tree can be used to create design rules: selecting a leaf node, the path from the root node to this leaf node is the decision making rule for a particular design. In this study, C4.5 algorithm [45] in WEKA, which is based on information entropy theory was adopted to build the decision tree.

In addition, the 3N experimental design method was used to determine the size of design space [35,46,47], where N is the number of design variables. Generally, the size of design space should be greater than 3N. A larger number of design alternatives can often improve the accuracy of modeling, however, it also increases the computational cost[35,46,47]. Based on the number of design variables, five (5) in this study and the other work in the literature [33,48,49], totally 150 LFT designs were created by Latin Hypercubic sampling algorithm to form the design space. The sample size should be sufficient to ensure the convergence of the results.

## 3. Results

### 3.1. The effects of filling Lattice structure

In order to analyze the effects of lattice structure filled in thin-walled tube on its EA performance, a comparison of SEA was conducted between LFT and corresponding hollow thin-walled tube with five thicknesses as shown in Figure 6. The SEAs of hollow thin-walled tubes with thickness t=0.8-, 1.1-, 1.4-, 1.7-, and 2.0- mm were 7.50-, 9.76-, 11.03-, 12.90-, and 13.64- kJ/kg, respectively. The result shows that in 120 out of 150 LFTs, the SEAs are higher than those of the hollow counterparts with the same thickness. Among these 120 LFTs, 70 (46.7% of the total designs) LFTs' SEA increased by over 20%, and 36 (24% of the total designs) LFTs' SEA increased by over 50%. The highest increase ratio of SEA was about 398.75%. This LFT had a 0.8 mm wall thickness tube and a lattice filler with $m=5$, $n=6$ and $d=3.0$ *mm*.

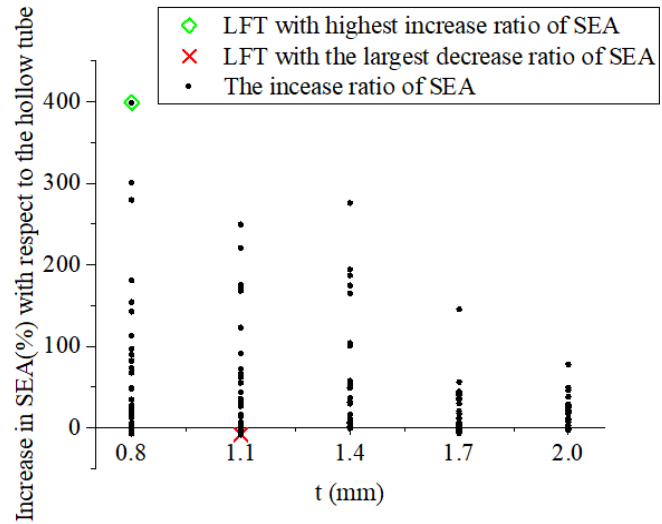

**Figure 6.** A comparison of SEAs of LFTs and corresponding hollow thin-walled tubes with the same thicknesses

The deformation pattern of this LFT is compared with that of the hollow thin-walled tube with a 0.8 mm thickness as shown in Figure 7a. It can be seen that the deformation of LFT exhibits a more uniform folding pattern compared with the hollow tube. Although, the SEAs of most LFTs were improved, 30 (20% of the total designs) LFTs' SEA decreased. The largest decrease ratio of SEA (-8.74%) occurred on the LFT with a 1.1 mm thickness tube and $m$ =4, $n$=3 and $d$=1.0 mm lattice structure. The deformation pattern of this LFT is also compared with that of the hollow tube with 1.1 mm thickness in Figure 7b. It is seen that the deformation pattern of LFT had no significant change after filling lattice structure. The increased mass due to filler led to the reduction in SEA of this LFT.

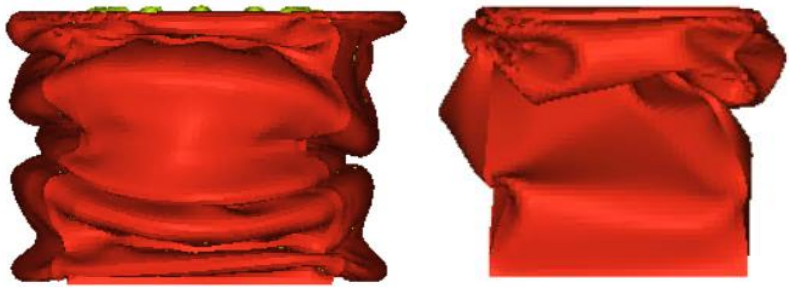

a. LFT with the highest increase ratio of SEA (left) and the corresponding hollow tube (right)

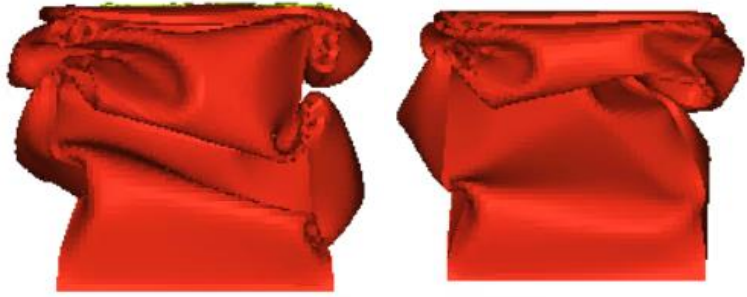

b. LFT with the largest decrease ratio of SEA (left) and the corresponding hollow tube (right)

**Figure 7.** Comparisons of deformation patterns of two LFTs and the corresponding hollow tubes with the same thicknesses

### 3.2. Univariate analysis

In order to analyze the effects of design variables on the SEA of LFTs, a statistical analysis for the SEAs of the 150 LFTs and their design variables was conducted and the results are shown in Figures 8a, to 8e, respectively.

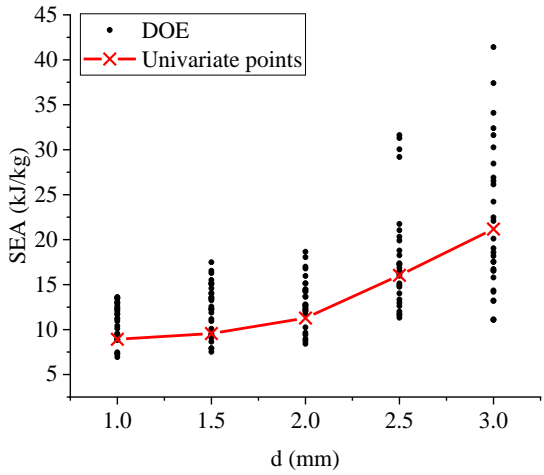

a. Diameter of lattice structure rod $d$

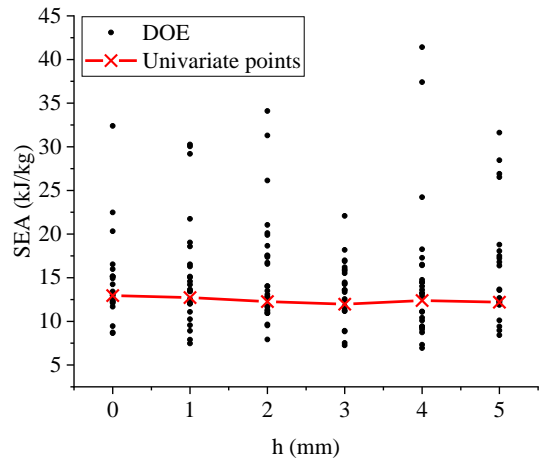

b. Height difference between lattice structure and tube $h$

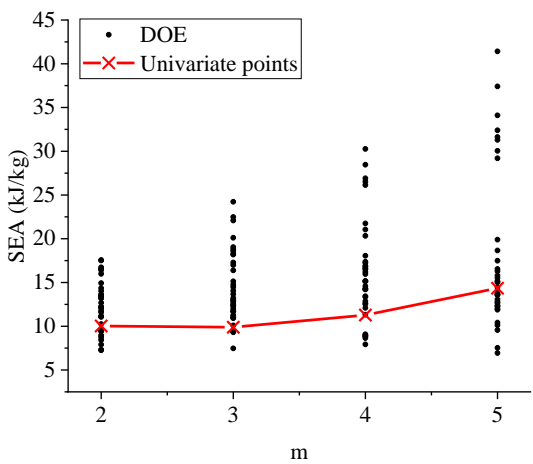

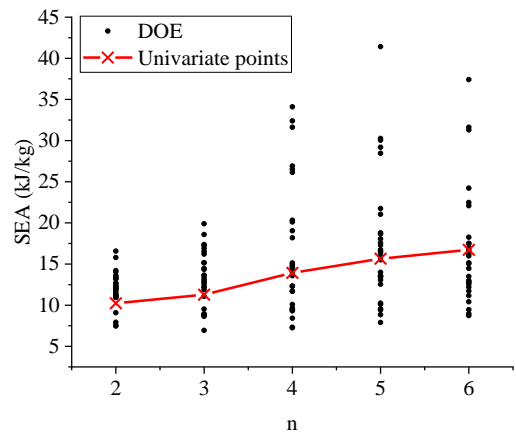

c. Number of cells in the transverse direction *m*    d. Number of layers in the longitudinal direction *n*

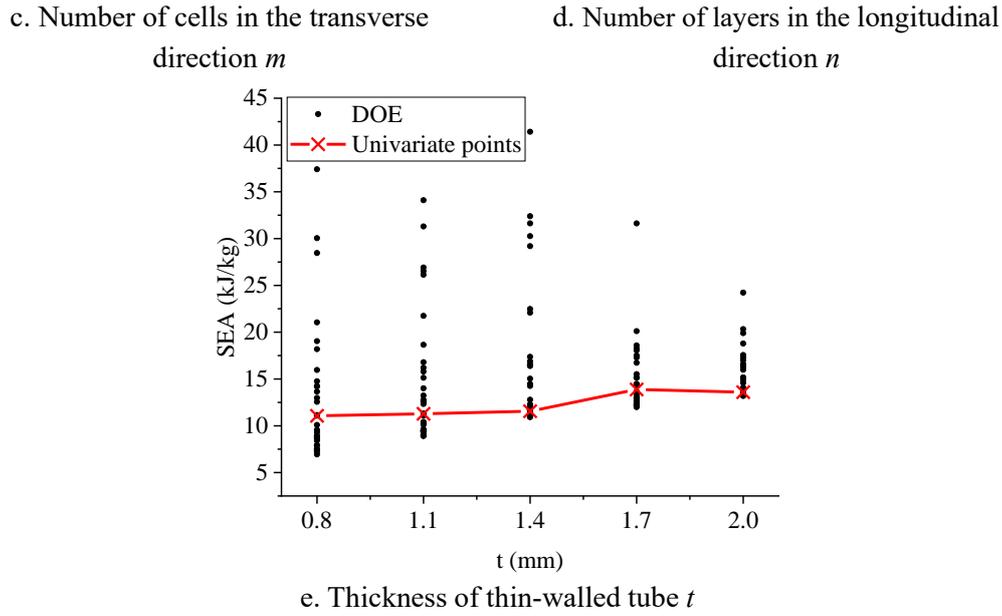

e. Thickness of thin-walled tube *t*

**Figure 8.** The results of statistical analysis for the effect of five design variables on SEA (At the univariate points from the same figure, the other four design variables remained as constants).

In Figure 8a, LFTs with higher SEA are mainly concentrated in the region with larger rod diameter *d*. When the other four design variables remained as constants, *m=4, n=3, t=1.1 mm*, and *h=3 mm*, respectively, the SEA of LFTs significantly increased with increase of *d*. Figure 8b shows that the distribution of the SEA of LFTs with different *h* values was scattered and had no clear trend. When the other four design variables were kept as constants, *m=4, n=3, t=1.1 mm*, and *d=2.0 mm*, respectively, the change of height difference *h* had little effect on SEA. In Figures 8c and 8d, the distributions of LFTs' SEA are similar as that in Figure 8a, *i.e.* more LFTs had higher SEA when *m* and *n* were larger. Also, increasing the values of *m* and *n* would improve *SEA*. In Figure 8e, LFTs with higher SEA were observed with *t*=1.1-, 1.4- and 1.7- mm. When the other four design variables remained as constants: *m=4, n=3, h=3 mm*, and *d=2.0 mm*, respectively, the change of tube thickness *t* did not significantly influence the SEA. However, when the tube thickness t was increased from 1.4 mm to 1.7 mm, SEA had an evident increase.

### 3.3. Data mining and validation of decision tree

In the univariate analysis, although the effect of each design variable on the SEA of LFT was obtained, it is still difficult to analyze the coupling effect of multiple design variables and also not possible to provide a guidance for design of LFT. Therefore, data mining based on the decision tree algorithm was employed to analyze the complex interrelations of design variables and derive the design rules. Three different design objectives, namely EA efficiency, total EA, and lightweight, were considered in data mining, respectively.

3.3.1. Decision tree on EA efficiency

To consider EA efficiency, SEA and CFE were used to classify LFTs in the simulation dataset. Based on the SEA and CFE values, the performance of LFTs was classified at three levels, i.e. Excellent/e, General/g, and Bad/b, respectively. The labeling criteria can be set in an arbitrary way based on the costumer's requirements or the designer's experience. According to a previous study on an energy absorber with the similar size[36], the criteria for classification for each level in this study were determined as follows:

1. Excellent/e: E1={SEA≥16 kJ/kg and CFE≥45%}
2. General/g: G1={SEA≥13.64 kJ/kg and CFE≥35%}-E1
3. Bad/b: L1=U-E1-G1

where U is the whole dataset (including 150 design alternatives) collected in this study; E1, G1 and L1 are the subsets of this dataset.

Using the above criteria for each level, a decision tree was generated by data mining though C4.5 algorithm (J48 in the Weka software) without any post-processing or simplification (i.e. "pruning" [50,51]) as shown in Figure 9. This decision tree includes 17 leaf nodes and 33 branches. The root node is the diameter $d$ of lattice structure rod, and the red path is one of the branches for "excellent/e" designs. The accuracies of "e", "g", and "b" classification which can be obtained by total number correctly classified DOEs divided total number DOEs under each classification are 91.3%, 73.5%, and 100%, respectively, and the average accuracy of whole decision tree is 88.3%. The accuracy of the partition for each leaf node can be indicated by the numbers on each leaf node. For example, on leaf node A in Figure 9, "e (13.0/2.0)" means that 13 samples were classified as "Excellent/e" designs with 2 non-"Excellent/e" design alternatives included in the prediction.

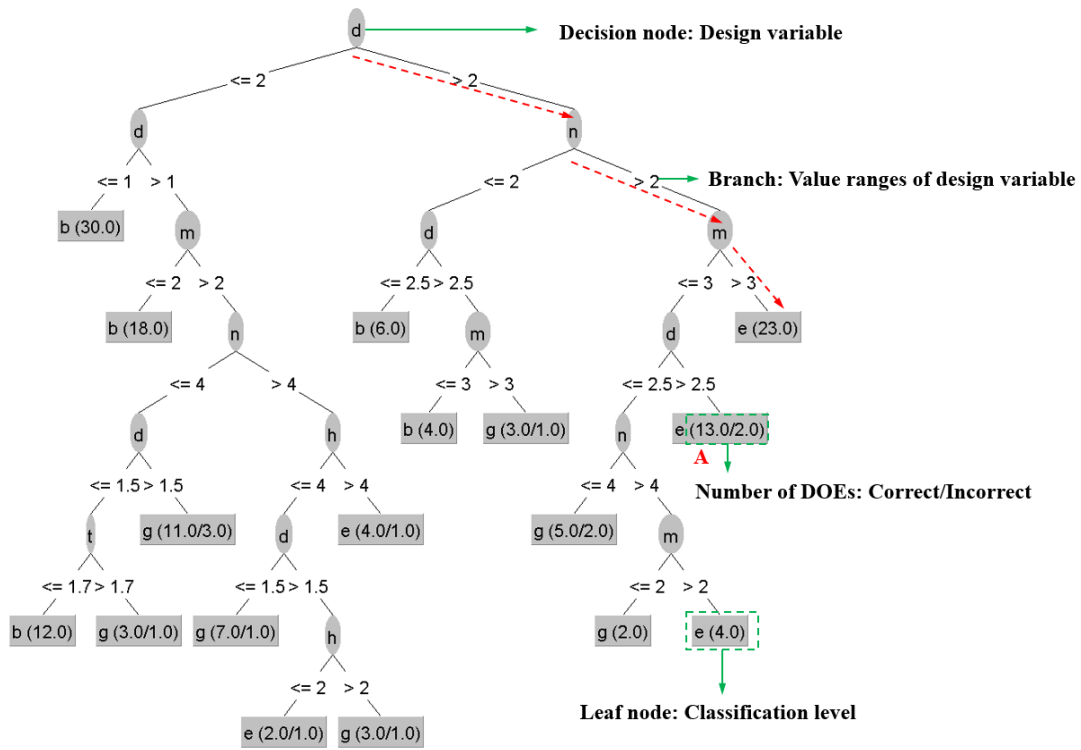

**Figure 9.** The decision tree on EA efficiency without pruning (Classification levels for LFT performance: e= "Excellent"; g= "General", and b= "Bad")

Although the classification accuracy of this decision tree is relatively high, there are too many branches, which indicates very complex design rules. Therefore, pruning is necessary to simplify the decision tree and decision making rules. Pessimistic-Error Pruning (PEP) proposed by Quinlan (1986)[51] is one of the Post-Pruning methods implemented for C4.5 in WEKA. Using PEP, the leaf nodes with low confidence on the original decision tree is then removed and their parent nodes or higher level parent nodes are replaced by new leaf nodes, which are those with highest probability to fall into this branch. Therefore, using PEP, an appropriate confidence factor should be defined as the threshold to check the confidence of leaf node and initiate the pruning process. In this study, it was required that the average accuracy of whole decision tree after pruning should be higher than 80%[36,38]. Based on this requirement, the original decision shown in Figure 9 was pruned, and the final decision tree on EA efficiency only consisted of 12 leaf nodes and 23 branches as shown in Figure 10. In this improved decision tree, the accuracies of "e", "g", and "b" are 93.2%、72.7% and 95.9%, respectively, and the average accuracy of whole decision tree is 87.3%.

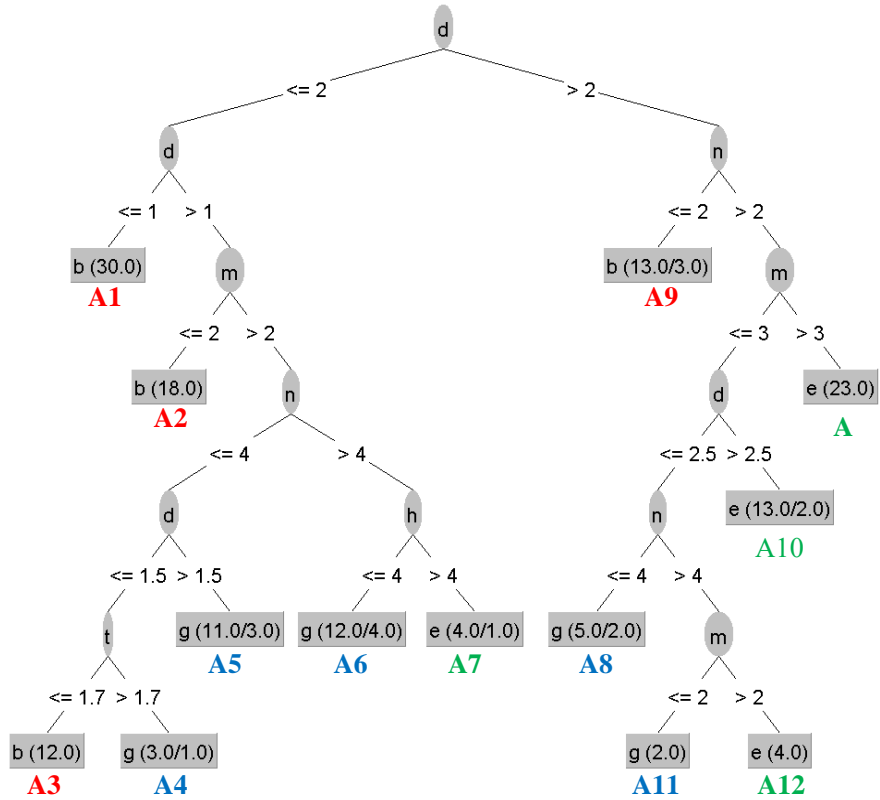

**Figure 10.** The decision tree on EA efficiency after pruning (Classification levels for LFT performance: e= "Excellent"; g= "General", and b= "Bad")

Figure 10 shows that the diameter $d$ of lattice structure rod is still the root node which indicates that it is the most influential variable and then followed by $n$ and $m$. In this decision tree, leaf nodes A, A7, A10, and A12 are classified as "e" level. Among these four paths, Branch A had the shortest path, which indicates a relatively simple decision making process[50]. It also should be noted that the classification accuracy of leaf node A is 100% (the number of incorrectly classified DOEs is zero). Therefore, the branch for leaf node A ($d>2$, $n>2$, and $m>3$) in Figure 10 was selected as the decision-making rule to design the LFT with the "e" level EA efficiency. i.e. let the design variables meet d>2 mm, n>2, and m>3, LFTs with EA efficiency "e" level (Excellent/e: SEA≥16 kJ/kg and CFE≥45%) can be achieved.

In order to validate the final decision tree on EA efficiency, a total of 15 new design alternatives were generated by following three different design rules from leaf nodes A, A5, and A1 which are respectively formed from the branches of "e", "g", and "b" leaf node with the highest accuracy. Under each design rule, 5 new design alternatives were created and simulated. The configurations and simulation results of these 15 new design alternatives are shown in Table 3.

In the "e" level design space, the simulation results for all five new design alternatives (i.e. actual labels) have demonstrated an "e" level performance on EA efficiency, which indicates that the accuracy of this design rule is 100%. The accuracies of "g" and "b" level design rules are about 60% and 86.7%, respectively, and the average accuracy of three design rules is 86.7%.

**Table 3.** Validation for the decision tree on EA efficiency

| NO | d/mm | n | m | h/mm | t/mm | SEA/(kJ/kg) | CFE/% | Actual label |
|---|---|---|---|---|---|---|---|---|
| "e" (SEA≥16 kJ/kg and CFE≥45%) design rule: d>2, n>2, m>3 (The branch for leaf node A) | | | | | | | | |
| 1 | 2.2 | 6 | 5 | 5 | 1.1 | 26.81 | 124.18 | e |
| 2 | 2.4 | 5 | 4 | 1 | 2.0 | 21.12 | 59.85 | e |
| 3 | 2.8 | 3 | 4 | 2 | 1.4 | 20.78 | 70.23 | e |
| 4 | 2.6 | 5 | 4 | 4 | 0.8 | 22.53 | 119.24 | e |
| 5 | 3.0 | 4 | 5 | 1 | 1.7 | 32.36 | 103.81 | e |
| "g" (13.64 kJ/kg≤SEA<16 kJ/kg and 35%≤CFE<45%) design rule: 1.5<d≤2, m>2, n≤4 (The branch for leaf node A5) | | | | | | | | |
| 1 | 1.6 | 4 | 5 | 5 | 1.1 | 14.07 | 47.95 | g |
| 2 | 1.7 | 3 | 3 | 1 | 2.0 | 14.23 | 35.69 | g |
| 3 | 1.9 | 2 | 4 | 2 | 1.4 | 10.56 | 30.23 | b |
| 4 | 1.8 | 3 | 4 | 4 | 0.8 | 10.96 | 38.87 | b |
| 5 | 2 | 3 | 5 | 1 | 1.7 | 14.67 | 40.86 | g |
| "b" (SEA<13.64 kJ/kg and CFE<35%) design rule: d≤1 (The branch for leaf node A1) | | | | | | | | |
| 1 | 1 | 6 | 5 | 0 | 2 | 13.62 | 31.45 | b |
| 2 | 1 | 2 | 3 | 1 | 1.7 | 12.8 | 28.14 | b |
| 3 | 1 | 4 | 4 | 4 | 0.8 | 8.09 | 20.94 | b |
| 4 | 1 | 3 | 4 | 2 | 1.4 | 11.07 | 25.96 | b |
| 5 | 1 | 5 | 2 | 5 | 1.1 | 9.73 | 21.67 | b |

3.3.2. Decision tree on total EA

In this section, another data mining was conducted on the same dataset in terms of the total EA, in additional to SEA. Therefore, there are two objectives, i.e. SEA and total EA (TEA). Based on their values, the performances of LFTs were classified at three levels and labeled as Excellent/e, General/g, and Bad/b. The criteria for each level[36] in this study are shown as follows:

1. Excellent/e: E2={SEA≥16 kJ/kg and TEA≥6 kJ}
2. General/g: G2={SEA≥13.64 kJ/kg and TEA≥14.45kJ}-E2
3. Bad/b: L2=U-E2-G2

where U is the whole dataset (including 150 design alternatives) collected in this study; E2, G2 and L2 are the subsets of this dataset.

C4.5 algorithm was used again to generate the second decision tree as shown in Figure 11. This decision tree consisted of 13 leaf nodes and 25 branches. The accuracies of "e", "g", and "b" are 90.7%、80% and 90.8%, respectively, and the average accuracy of this whole decision tree is 87.2%. The shortest branches for the "e" leaf nodes (Leaf nodes B) was selected as the design rules. For "e" level designs, d>2 mm, n>2, and m>3.

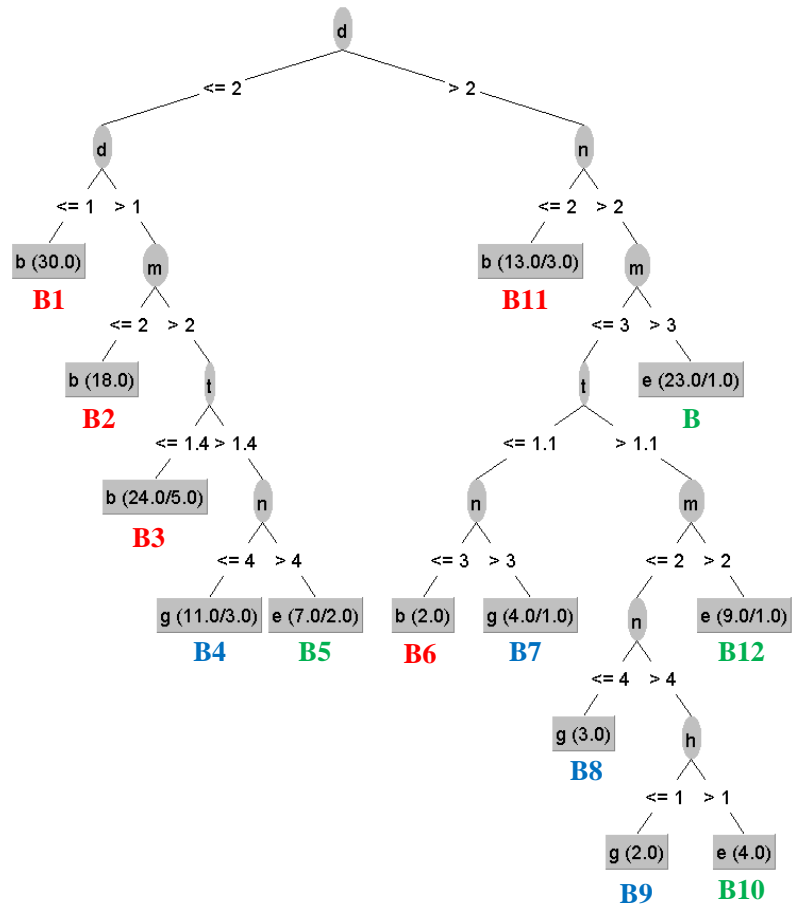

**Figure 11.** The decision tree on total EA after pruning

The decision tree shown in Figure 11 was validated in the same way as described in Section 3.3.1 and the results are summarized in Table 4. The accuracies of "e", "g", and "b" level design rules are about 100%, 40%, and 80%, respectively, and the average accuracy of three design rules is 80.0%.

**Table 4.** Validation for the decision tree on total EA

| NO | d/mm | n | m | h/mm | t/mm | SEA/(kJ/kg) | EA/kJ | Actual label |
|---|---|---|---|---|---|---|---|---|
| "e" (SEA≥16 kJ/kg and EA≥6kJ) design rule: d>2, n>2, m>3 (The branch for leaf node B) | | | | | | | | |
| 1 | 2.2 | 6 | 5 | 5 | 1.1 | 26.74 | 10.61 | e |
| 2 | 2.4 | 5 | 4 | 1 | 2.0 | 20.63 | 10.28 | e |
| 3 | 2.8 | 3 | 4 | 2 | 1.4 | 21.21 | 9.32 | e |
| 4 | 2.6 | 5 | 4 | 4 | 0.8 | 23.24 | 7.59 | e |
| 5 | 3.0 | 4 | 5 | 1 | 1.7 | 33.69 | 21.86 | e |
| "g" (13.64 kJ/kg≤SEA<16 kJ/kg and 4.45kJ≤EA<6kJ) design rule: 1<d≤2, m>2, 2<n≤4, t>1.4 (The branch for leaf node B4) | | | | | | | | |
| 1 | 1.2 | 4 | 5 | 5 | 1.6 | 11.64 | 3.82 | b |
| 2 | 1.4 | 3 | 3 | 1 | 2 | 14.11 | 5.13 | g |
| 3 | 1.8 | 2 | 4 | 2 | 1.8 | 12.03 | 4.56 | b |
| 4 | 1.6 | 3 | 4 | 4 | 1.8 | 13.07 | 4.78 | b |
| 5 | 2 | 3 | 5 | 1 | 2 | 15.21 | 7.66 | g |
| "b" (SEA<13.64 kJ/kg and EA<4.45) design rule: d≤1 (The branch for leaf node B1) | | | | | | | | |
| 1 | 1.2 | 6 | 2 | 5 | 2 | 13.11 | 4.55 | b |
| 2 | 1.4 | 2 | 2 | 1 | 1.7 | 12.78 | 3.77 | b |
| 3 | 1.8 | 4 | 2 | 3 | 0.8 | 8.25 | 1.33 | b |
| 4 | 1.6 | 3 | 2 | 4 | 1.4 | 10.77 | 2.70 | b |
| 5 | 2 | 5 | 2 | 1 | 1.1 | 9.88 | 2.17 | b |

3.3.3. Decision tree on the structural weight

Another decision tree was created to implement the lightweight design of LFT, in addition to SEA. The three levels of the structural lightweight design were defined using the following

criterion[36]:

1. Excellent/e: E3={SEA≥16 kJ/kg and Mass≤0.45kg}
2. General/g: G3={SEA≥13.64 kJ/kg and Mass≤0.5kg}-E3
3. Bad/b: L3=U-E3-G3

where U is the whole dataset (including 150 design alternatives) collected in this study; E3, G3 and L3 are the subsets of this dataset.

The same algorithm was used to build the decision tree which consisted of 12 leaf nodes and 25 branches as shown in Figure 12. The accuracies of "e", "g", and "b" labeling are 78.8%, 73.7% and 88.8%, respectively, and the average accuracy of this whole decision tree is 80.42%. The shortest branch for "e" class (Leaf nodes C) was selected as the design rule. For "e" level designs, d>2mm, n>2, m≤4, and t≤1.7 mm.

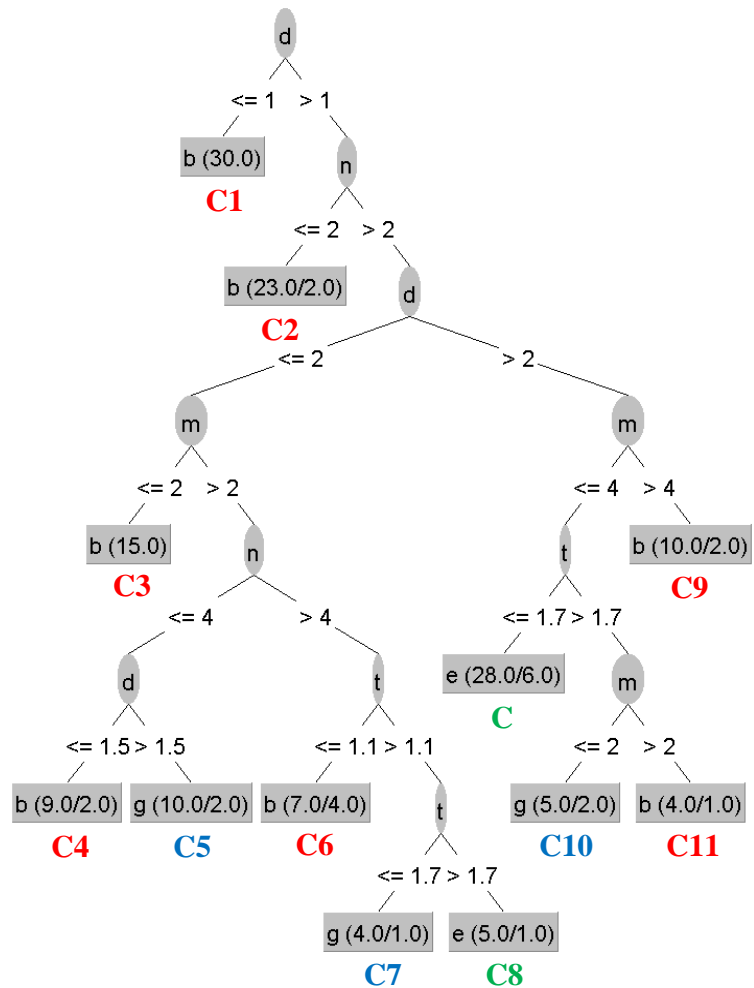

**Figure 12.** The decision tree on lightweight after pruning

The validation was conducted in the same way as described above and the results are shown in

Table 5. The accuracies of "e" and "b" level design rules are about 80% and 100%, respectively, and the average accuracy of three design rules is 73.3%. It also verified that the accuracy of decision tree on lightweight in Figure 12 is sufficient to design LFTs with "e" level lightweight (SEA≥16 kJ/kg and Mass≤0.45kg). It should be noted that the accuracy of "g" level design rules is lower than that of "e" and "b" level design rules in all three decision trees and all cases for validations. This is probably due to the fact that the design space for "g" class is relatively small and its design rules are more complicated. More design alternatives at this level are needed to improve its predication accuracy.

**Table 5.** Validation for the decision tree on lightweight

| NO | d/mm | n | m | h/mm | t/mm | SEA/(kJ/kg) | Mass/kg | Actual label |
|---|---|---|---|---|---|---|---|---|
| "e" (SEA≥16 kJ/kg and Mass≤0.45) design rule: d>2, n>2, m≤4, t≤1.7 (The branch for leaf node C) | | | | | | | | |
| 1 | 2.2 | 6 | 4 | 5 | 1.1 | 20.35 | 0.33 | e |
| 2 | 2.4 | 5 | 2 | 1 | 1.7 | 13.56 | 0.34 | b |
| 3 | 2.8 | 3 | 3 | 2 | 1.4 | 16.09 | 0.36 | e |
| 4 | 2.6 | 5 | 3 | 4 | 0.8 | 16.03 | 0.25 | e |
| 5 | 3.0 | 4 | 4 | 1 | 1.1 | 27.50 | 0.43 | e |
| "g" (13.64 kJ/kg≤SEA<16 kJ/kg and 0.45<Mass≤0.5) design rule: 1.5<d≤2, m>2, 2<n≤4 (The branch for leaf node C5) | | | | | | | | |
| 1 | 1.6 | 4 | 5 | 5 | 1.1 | 12.89 | 0.29 | b |
| 2 | 1.7 | 4 | 3 | 1 | 2.0 | 14.34 | 0.38 | g |
| 3 | 1.9 | 3 | 4 | 2 | 1.4 | 13.04 | 0.33 | b |
| 4 | 1.8 | 4 | 4 | 4 | 0.8 | 13.32 | 0.22 | b |
| 5 | 2.0 | 3 | 5 | 1 | 1.7 | 14.39 | 0.45 | g |
| "b" (SEA<13.64 kJ/kg and Mass>0.5) design rule: d>1, n≤2 (The branch for leaf node C2) | | | | | | | | |
| 1 | 1.2 | 2 | 5 | 5 | 2 | 13.58 | 0.38 | b |
| 2 | 1.6 | 2 | 2 | 1 | 1.7 | 12.08 | 0.30 | b |
| 3 | 2.4 | 2 | 4 | 3 | 0.8 | 10.38 | 0.27 | b |
| 4 | 2.0 | 2 | 3 | 4 | 1.4 | 11.20 | 0.29 | b |
| 5 | 2.8 | 2 | 4 | 1 | 1.1 | 13.07 | 0.37 | b |

## 4. Discussions

In the traditional parametric analysis presented in Sections 3.1 and 3.2, the relationships between each individual variable and the SEA of LFT were determined. However, the decision-making rules to design a LFT with desired performance were still not available. Therefore, the data mining approach was used to obtain the coupling effect between design variables in the form of a decision tree. In Sections 3.3.1, 3.3.2, and 3.3.3, three decision trees were established based on the different design requirements (EA efficiency, total EA and lightweight), respectively.

Typical designs at all three classification levels in terms of all three design objectives and their deformation patterns are compared in Table 6. The results showed that regardless of the design objectives, the LFTs in the same classification level have similar configuration, i.e. the designs with "e" labels have higher $m$ and $n$ values than those with "g" and "b" labels. In other words, the "e" group lattice structure has a higher density. Zhang et al. (2015)[52] and Maskery et al. (2016)[53] also reported that optimization of lattice structure's density could better organize the distribution of materials and improve the utilization efficiency. Therefore, the density of lattice structure should be taken into consideration in the design of similar structures. Furthermore, compared with the LFTs in the "g" or "b" classes, the deformation patterns of LFTs in the "e" group are more uniform to absorb higher energy.

**Table 6.** The LFTs and their deformation patterns in the three classes

| Objective | SEA and CFE | SEA and EA | SEA and Mass |
|---|---|---|---|
| Level "e" | 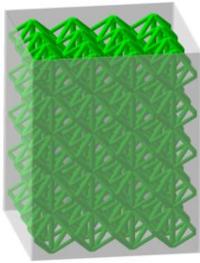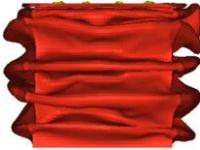 | 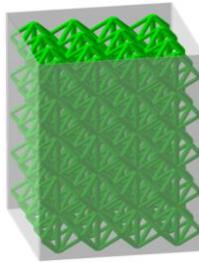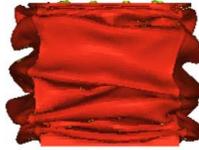 | 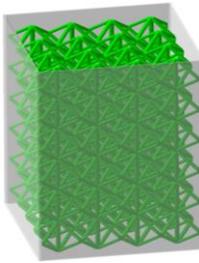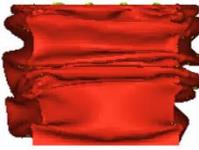 |

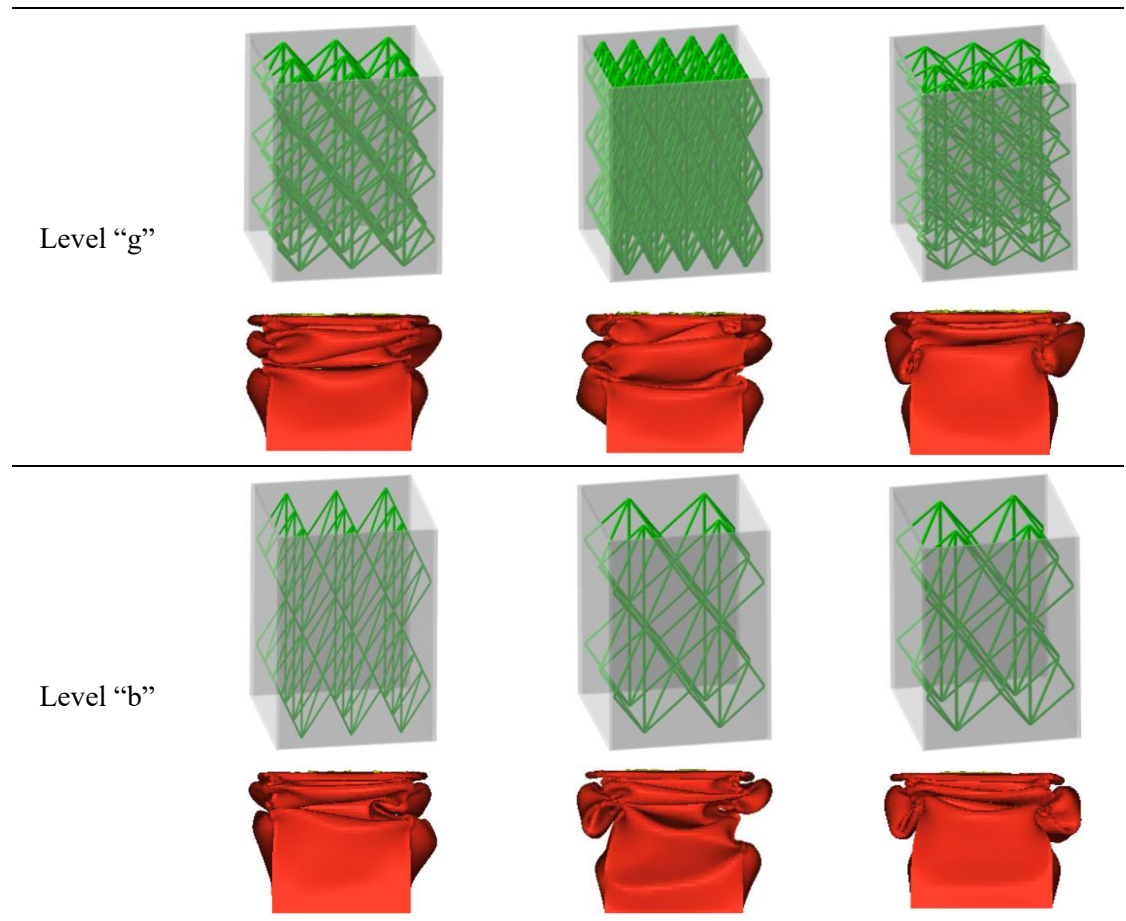

| | | | |
|---|---|---|---|
| Level "g" | | | |
| Level "b" | | | |

In summary, these decision trees are not able to directly produce global optimal design. However, the design rules derived can be used to generate a group of designs or a design set satisfying the design requirements. Compare to the optimization approach, this method is more robust and has a larger tolerance to the uncertainties in the simulations and manufacturing. Therefore, it is more practical to use in the actual engineering design. A detailed uncertainty analysis associated with this method can be seen in Du's Doctorate Dissertation[54].

## 5. Conclusions

A new type of energy absorbing structure, named as lattice-structure filled thin-walled tube (LFT), was studied and designed based on data mining-decision tree algorithm in this paper. In the 3-D parametric design of LFT, five design variables, namely number of layers in the longitudinal direction $n$, number of cells in the transverse direction $m$, rod diameter $d$, tube thickness $t$, and height difference between tube and lattice structure $h$, have been considered. Using Latin Hypercubic sampling algorithm, 150 design alternatives were generated and simulated in an automation platform. In this design process, the crashworthiness indices, including *TEA*, *SEA*, $P_m$, *CFE* etc. were calculated and collected from simulation results to form a dataset for data mining. The traditional parametric analysis was conducted first and the results showed that: (1). Filling the BCC-Z (Body-centered cubic unit cell with vertical struts) lattice structure into the hollow thin-walled tube can significantly improve the EA capacity of the structure. (2). SEA of LFT increased by increasing the *d*, *m*, and *n*. To address the limitations of traditional parametric study, a decision tree algorithm (C4.5) was used to mine the dataset and create three decision trees and then derive corresponding design rules in terms of EA efficiency, total EA and lightweight requirements, respectively. Based on the design rules, the configurations of LFTs with "e" (i.e. excellent) label are determined as follows: For high EA efficiency (SEA≥16 kJ/kg and CFE≥45%), d>2 mm, n>2, and m>3; For high total EA (SEA≥16 kJ/kg and EA≥6kJ), d>2mm, n>2, and m>3; and For lightweight (SEA≥16 kJ/kg and Mass≤0.45kg), d>2 mm, n>2, m≤4, and t≤1.7 mm. This study demonstrates that the decision tree method is a powerful tool to interpret and obtain the structural design rules. Such rules can be used to generate new designs with satisfied performance without running additional simulations.

**Acknowledgement**

The authors at the Hunan University would like to thank the financial support from Innovative Research Groups of National Natural Science Foundation of China (Grant No. 51621004), National Natural Science Foundation of China (Grant No. 51475153 and 51405148).

**Data Availability Statement**

The raw/processed data required to reproduce these findings cannot be shared at this time due to technical or time limitations.